\documentstyle[aps,preprint]{revtex}

\begin{document}
\draft

\title{Statistical properties and  statistical interaction for particles with
spin:
Hubbard model in one dimension and\\
 statistical spin liquid}

\author{Krzysztof Byczuk$^{a}$ and Jozef Spa\l ek$^{b,a}$}

\address{(a) Institute of Theoretical Physics, Warsaw University,
ul. Ho\.za 69, 00-681 Warszawa, Poland \\
(b) Institute of Physics, Jagellonian University, ul. Reymonta 4, 30-059
Krak\'ow, Poland }

\maketitle

\begin{abstract}
We derive the statistical distribution functions for the Hubbard chain with
infinite Coulomb repulsion among particles and for the statistical spin
liquid with an arbitrary magnitude of the local interaction in  momentum
space.
Haldane's statistical interaction is derived  from an exact solution for each
of the two models.
In the case of the Hubbard chain the charge (holon) and the spin (spinon)
excitations decouple completely and are shown to behave statistically
as fermions and bosons, respectively.
In both cases the statistical interaction must contain several
components, a rule for the particles with the internal symmetry.
\end{abstract}

\newpage

\section{Introduction}

It is  well known  that in the space of dimension higher than two the
many-particle wave function is either
symmetric or antisymmetric under a permutation
group operation; this property leads to the division into the systems of
bosons and fermions, respectively.
As a consequence, the distribution function for the  ideal gas
is given either by
the Bose-Einstein or by the Fermi-Dirac functions \cite{hei}.
In low-dimensional systems ($d=1$ and $2$)
the situation changes drastically because
e.g., a proper
symmetry group in  two dimensions for the hard-core particles
is the  braid group, the  characters
of which are complex numbers \cite{l}.
In such instances the distribution function has not been determined
as yet.
On the other hand,  the distribution function can  be changed by
 the interaction among particles.
Such a situation arises for instance
 at the critical point when the system undergoes
a phase transition.
Below the  critical temperature (e.g., in the superconducting phase)
the distribution function changes its
form from that in the normal state.
So, the statistical properties of the particles are influenced by both system
dimensionality and by the character of dynamical interaction between particles.

In his remarkable paper \cite{hal}, Haldane noted that the distribution
function can  also differ from the Bose-Einstein or the
Fermi-Dirac form in
the normal state.
He generalized the Pauli exclusion principle by introducing
the concept of {\bf statistical interaction}
  which determines how the number of
accessible orbitals changes when   particles are added to the system.
The paper dealt with the many-particle Hilbert space of finite dimension.
The limitation turned out to be irrelevant.
Namely, Murthy and Shankar showed \cite{ms}
that the statistical interaction,
when extended to the  Hilbert space of infinite dimension, is proportional
to the second virial coefficient.

Very recently, Wu \cite{wu} solved
 the problem of the distribution function
for   Haldane's fractional
statistics.
He found a general form of the equations for the distribution function
for  an arbitrary
 statistical interaction and discussed the thermodynamics of such a gas.
Furthermore,
Bernard and Wu \cite{be} found the explicit form of the statistical
interaction in the case of  interacting scalar particles in one dimension.
In the particular case of  bosons  they showed that,
as the
amplitude of a local delta-function interaction changes from zero to
infinity,
the distribution function evolves
from the Bose-Einstein to the Fermi-Dirac form.

In this paper we introduce the spin degrees of freedom into the problem
and determine  the statistical properties as well as the statistical
interaction
for particles in two situations.
We consider  first the Hubbard model in  the space of one  dimension and with
an infinite on-site Coulomb repulsion.
In this limit, we show rigorously that the charge excitations (holons)
obey the Fermi-Dirac
distribution,
 whereas the spin excitations (spinons) obey the Bose-Einstein distribution.
 The boson part leads to the correct entropy ($k_B \ln 2$ per carrier) in the
 Mott insulating limit.
As  a second example,
 we express the statistical spin liquid partition function
\cite{s1} with the help of the statistical interaction concept.
These two examples represent
 nontrivial generalizations of  Haldane's fractional
statistics to particles with  internal symmetry such as spin.
In both cases an explicit form
of  the multicomponent statistical interaction is required.
We show that the statistical distributions evolve to some novel functions
when the interaction between the particles  diverges.
For the spin liquid case the form of the distribution functions are also
presented for  intermediate values of the dynamical interaction.
In both cases the nonstandard statistics is due to the interaction between
the particles.

\section{Statistical interaction for the Hubbard model in one dimension}
\subsection{Thermodynamic limit for the Bethe ansatz equations
\mbox{$(U \rightarrow \infty)$} }

We consider first the one-dimensional system of particles with
a contact interaction.
One of the simplest models
 of interacting spin one-half particles was introduced by
Hubbard \cite{hub}.
The Hamiltonian in this case is
\begin{equation}
H = - t \sum_{<i,j> \sigma} a_{i \sigma}^{+} a _{j \sigma}
+ U \sum_{i} n_{i \uparrow} n_{i \downarrow},
\end{equation}
where $t$ is the hopping integral between the nearest neighboring pairs
$<i,j>$ of lattice sites,  and $U$ is the on-site
Coulomb repulsion when the two particles with spin up
and down meet on the same lattice site.
We set $t=1$.
This model was solved in one dimension by Lieb and Wu \cite{lieb}.
The solution is given by the set of the Bethe-ansatz equations
determining the rapidities $\{k_i\}$, $\{ \Lambda_{\alpha}\}$; i.e.,
\begin{equation}
\renewcommand{\arraystretch}{1.5}
\everymath={\displaystyle}
\left\{
\begin{array}{c}
\frac{2 \pi}{L} I_{j} = k_{j} - \frac{1}{L} \sum_{\beta =1}^{M}
\Theta ( 2 \sin k_{j} - 2 \Lambda_{\beta} ) \\
\frac{2 \pi}{L} J_{\alpha} = \Lambda_{\alpha} - \frac{1}{L}
\sum_{j = 1}^{N} \Theta ( 2 \Lambda_{\alpha} - 2 \sin k_{j} ) +
\frac{1}{L} \sum_{\beta =1 }^{M} \Theta( \Lambda _{\alpha} -
\Lambda_{\beta})
-  \sum_{\beta =1}^{M} \Lambda_{\beta}
\delta_{\Lambda_{\alpha},\Lambda_{\beta}} ,
\end{array}
\right.
\end{equation}
where $N$ is the total number of particles in the system, $M$ is the number
of particles with spin down, $L$ is the length of the chain, $j=1,...,N$,
and
$\alpha=1,...,M$.
$I_{j}$ is an integer (half-odd integer) for M even (odd), and
$J_{\alpha}$ is an integer (half-odd integer) for $N-M$ odd (even).
The phase shift function $\Theta (p)$ is defined by
\begin{equation}
\Theta (p) = - 2 \tan ^{-1} \left( \frac{2 p}{U} \right).
\end{equation}
The second set of equations (2) was written in the form better
suited to our purposes.
The basis in the Hilbert space which diagonalizes the Hamiltonian (1)
is called the holon-spinon representation.

The
Bethe-ansatz equations can be rewritten in such a way that all  dynamical
interactions are transmuted into the statistical interaction \cite{be}.
We determine explicitly the
 statistical interaction in the case of the  Hubbard model.
Our method is a straightforward generalization of the Bernard and Wu result
\cite{be} and is valid in the case of infinite interaction only.
In this limit, the charge and the spin excitations decouple  and
there are no bound states in the system \cite{ogata}.
In other words, all bound states in the upper Hubbard subband are pushed
out from the physical many-particle Hilbert space.

In the large $U$ limit the Bethe ansatz equations read \cite{sok}
\begin{equation}
\renewcommand{\arraystretch}{1.5}
\everymath={\displaystyle}
\left\{
\begin{array}{c}
\frac{2 \pi}{L} I_{j} = k_{j} + \frac{1}{L} \sum_{\beta =1}^{M}
\Theta ( 2 \Lambda_{\beta} ) \\
\frac{2 \pi}{L} J_{\alpha} = \Lambda_{\alpha} - \frac{N}{L}
\Theta ( 2 \Lambda_{\alpha} ) +
\frac{1}{L} \sum_{\beta =1 }^{M} \Theta( \Lambda _{\alpha} -
\Lambda_{\beta})
-  \sum_{\beta =1}^{M} \Lambda_{\beta}
\delta_{\Lambda_{\alpha}, \Lambda_{\beta}} .
\end{array}
\right.
\end{equation}
We rewrite these equations in the
thermodynamic limit, i.e. for  $N\rightarrow \infty$, $L\rightarrow \infty$,
and $N/L = const$.
We divide the range of the momentum $k's$ and $\Lambda 's$ into the
 intervals
with an equal size $\Delta k$ and $\Delta \Lambda$, as well as
label each interval
by its midpoints $k_i$ and $\Lambda _{\alpha}$, respectively.
We treat the particles with the momenta in the $i$-th or the $\alpha$-th
interval as belonging to the $i$-th or the  $\alpha$-th group.
As usual, the number of available bare single particle states are
$
G_{i}^{0} = L \Delta k/2 \pi
$ and
$
G_{\alpha}^{0} = L \Delta \Lambda/ 2 \pi.
$
These numbers follow from the decomposition of the Bethe-ansatz wave function
in the $U \rightarrow \infty$ limit \cite{ogata}.
Next, we introduce the distribution functions (the densities of states)
for the roots $k_{j}$ and $\Lambda_{\alpha}$ of the Bethe-ansatz
equations (4).
Namely, we define
$
L\rho (k_i) \Delta k \equiv N_{i}^{c}
$
as the number of $k $ values in
the interval $ [ k_i - \Delta k /2, k_i + \Delta k/2 ]$,
and
$
L\sigma(\Lambda_{\beta})\Delta\Lambda \equiv N_{\beta}^{s}
$
as the number of
$\Lambda$ values in the interval
 $[ \Lambda_{\beta} - \Delta \Lambda /2, \Lambda_{
\beta} + \Delta \Lambda /2 ]$.
Hence, the two quantities
$
2 \pi \rho (k_i ) = N_{i}^{c}/G_{i}^{0} \equiv n_{i}^{c}
$,
and
$
2 \pi \sigma (\Lambda_{\beta}) = N_{\beta}^{s}/G_{\beta}^{0}
\equiv n_{\beta}^{s}
$,
are respectively the occupation-number distributions
for the holon and the spinon excitations
in the Hubbard chain.
In effect, the Bethe
ansatz equations in the intervals $\Delta k$ and $\Delta \Lambda$
take the form
\begin{equation}
\renewcommand{\arraystretch}{1.5}
\everymath={\displaystyle}
\left\{
\begin{array}{c}
\frac{2 \pi}{L} I(k_{j}) = k_{j} +  \sum_{\beta}
\Theta ( 2 \Lambda_{\beta} ) \sigma(\Lambda_{\beta}) \Delta \Lambda \\
\frac{2 \pi}{L} J(\Lambda_{\alpha}) = \Lambda_{\alpha} -
\frac{N}{L} \Theta ( 2 \Lambda_{\alpha} ) +
\sum_{\beta } \Theta( \Lambda _{\alpha}  -
\Lambda_{\beta}) \sigma(\Lambda_{\beta}) \Delta \Lambda\\
 - L \sum_{\beta} \Lambda_{\beta}
\delta_{\Lambda_{\alpha},\Lambda_{\beta}} \sigma( \Lambda_{\beta}) \Delta
\Lambda.
\end{array}
\right.
\end{equation}
The function $\rho(k)$ does not appear explicitly in the large $U$ limit.

The numbers of accessible states in each of the $i$-th and
the $\alpha$-th groups
are
\begin{equation}
\tilde{D}_{i}^{c} (\{N_{i}^{c}\},\{N_{\beta}^{s}\}) =
I(k_i + \Delta k /2) - I (k_i - \Delta k /2),
\end{equation}
\begin{equation}
\tilde{D}_{\alpha}^{s} (\{N_{i}^{c}\},\{N_{\beta}^{s}\})=
J( \Lambda _{\alpha} + \Delta \Lambda /2 ) - J (\Lambda _{\alpha} - \Delta
\Lambda /2).
\end{equation}
Using the continuous  form (5) of the Bethe ansatz equations
 we find that
$
\tilde{D}_{i}^{c} = L \rho _{t}^{c}(k_{i}) \Delta k $
and
$
\tilde{D}_{\alpha}^{s} = L \rho _{t}^{s} (\Lambda_{\alpha}) \Delta \Lambda,
$
where in the thermodynamic limit  ($\Delta k \rightarrow 0$ , $\Delta
\Lambda \rightarrow 0$)  we have respectively
the total densities of states
for charge and spin excitations
\begin{equation}
\rho_{t}^{c} (k) = \frac{1}{2 \pi},
\end{equation}
and \cite{got}
\begin{equation}
\rho_{t}^{s} (\Lambda) = \frac{1}{2 \pi} - \frac{N}{2 \pi L}
\frac{\partial \Theta(2 \Lambda)}{\partial \Lambda} +
\frac{1}{2 \pi}\int d\Lambda' \sigma(\Lambda ') \frac{\partial \Theta(
\Lambda - \Lambda ')}{\partial \Lambda}-
 \int d \Lambda ' \sigma( \Lambda ') \Lambda '
\frac{\partial \delta( \Lambda - \Lambda ')}{\partial \Lambda}.
\end{equation}
Substituting the form (3) for $\Theta (p)$ to the derivative $
\partial \Theta / \partial \Lambda$ one can easily find that in the
$U\rightarrow \infty $ limit
\begin{equation}
\rho^{s}_{t}(\Lambda) =
\frac{1}{2 \pi}  + \sigma ( \Lambda).
\end{equation}
To derive (10) we utilized the fact that $\sigma (\Lambda)$ is a flat
function of $\Lambda$ in the large $U$ limit \cite{car}.
We see that the numbers of accessible states for the holons and the spinons
in the $U\rightarrow \infty$ limit are independent of each other.
This result, as we show in the following, leads to the decomposition
of the partition function into the holon and the spinon parts.

\subsection{Statistical interaction for the Hubbard chain}

We define the statistical interaction and the total number of
states for spin one-half particles.
For that purpose we work in the basis  in which the
Hamiltonian is diagonal, i.e.,
we choose the holon-spinon representation and
 observe that the dimensions $D_{i}^{c}$ and $D_{\alpha}^{s}$ of the
one-particle Hilbert spaces for the particle in the $i$-th or the $\alpha$-th
groups are functionals of both $\{N_{i}^{c}\}$ and $\{N_{\alpha}^{s}\}$, i.e.
$
D_{i}^{c} = D_{i}^{c} (\{N_{i}^{c}\},\{N_{\beta}^{s}\}),
$ and
$
D_{\alpha}^{s}=D_{\alpha}^{s}(\{N_{i}^{c}\},\{N_{\beta}^{s}\}).
$
Namely, starting from the Haldane definition \cite{hal} of the change of the
number of the accessible states and adopting it to the present situation we
obtain
\begin{equation}
\Delta D_{i}^{c} = -
 \sum_{j} g_{ij}^{cc} \Delta N_{j}^{c}
 - \sum_{\alpha} g_{i\alpha}^{cs} \Delta N_{\alpha}^{s},
\end{equation}
\begin{equation}
\Delta D_{\alpha}^{s} =  -
\sum_{j} g_{\alpha j}^{sc} \Delta N_{j}^{c} -
\sum_{\beta} g_{\alpha \beta}^{ss}
\Delta N_{\beta}^{s},
\end{equation}
where the four $g$ parameters are called the {\bf statistical interactions}.
These difference equations can be transformed to the following differential
forms:
\begin{equation}
(-g_{ij}^{cc})^{-1} \frac{\partial D_{i}^{c}}{\partial N_{j}^{c}} +
(-g_{i \alpha}^{cs})^{-1} \frac{\partial D_{i}^{c}}{\partial N_{\alpha}^{s}}
=2,
\end{equation}
\begin{equation}
(-g_{\alpha i }^{sc})^{-1} \frac{\partial D_{\alpha}^{s}}{\partial N_{i}^{c}} +
(-g_{\alpha \beta}^{ss})^{-1}
\frac{\partial D_{\alpha}^{s}}{\partial N_{\beta}^{s}}
=2.
\end{equation}
This set of equations establishes the generalization of Haldane's
equations for the number of accessible orbitals of the species
$\alpha$ in the case of particles without internal symmetry.
As before \cite{hal}, statistical interactions $\{g\}$ do not depend on the
occupations ${N_{\alpha}^s}$ and  ${N_i^c}$, since otherwise the thermodynamic
limit would not be well defined.
The factors $2$ in the r.h.s. of (13) and (14) are irrelevant because they can
be incorporated into the $g$ parameters.
Then the solutions of equations (13) and (14) are
\begin{equation}
 D_{i}^{c} (\{N_{i}^{c}\},\{N_{\beta}^{s}\}) = G_{i}^{0} -
 \sum_{j} g_{ij}^{cc} N_{j}^{c} - \sum_{\alpha} g_{i\alpha}^{cs}N_{\alpha}^{s},
\end{equation}
\begin{equation}
D_{\alpha}^{s}(\{N_{i}^{c}\},\{N_{\beta}^{s}\}) = G_{\alpha}^{0} -
\sum_{j} g_{\alpha j}^{sc} N_{j}^{c} - \sum_{\beta} g_{\alpha \beta}^{ss}
N_{\beta}^{s}.
\end{equation}
One should note that these solutions are well defined also in the boson limit,
since then the corresponding $g$ parameter(s) vanish.
The relations
$
D_{i}^{c}(\{0\},\{0\}) = G_{i}^{0}
$ and
$
D_{\alpha}^{s}(\{0\},\{0\}) = G_{\alpha}^{0}
$ express  the boundary conditions for this problem;  the values
$G_{\alpha}^s$ and $G_i^c$ represent the maximal values of available
one-particle states in the situation when the holon and the spinon
bands are empty.

Additionally,  the total number of microscopic
configurations with the numbers $\{N_{j}^{c}\}$ and
$\{N_{\beta}^{s}\}$  of holon and spinon excitations is given by
\begin{equation}
\Omega = \prod_{i=1}^{N} \frac{ (D_{i}^{c}+N_{i}^{c} -1)!}{(N_{i}^{c})!
(D_{i}^{c} -1)!} \prod_{\alpha=1}^{M}
\frac{(D_{\alpha}^{s} + N_{\alpha}^{s} -1)!}{(N_{\alpha}^{s})!(D_{\alpha}^{s}-
1)!}.
\end{equation}
In this expression the two products are in general interconnected via
the relations (15) and (16).
Each of the factors is defined in the same manner as in Ref.\cite{hal}.
 In the fermionic bookkeeping for $I_j$ and $J_{\alpha}$ the same $\Omega$
 is obtained with the number of accessible states in the $i$-th and
 $\alpha$-th groups taken to be \cite{be}
\begin{equation}
\tilde{D}_{i}^{c} (\{N_{i}^{c}\},\{N_{\beta}^{s}\})=
D_{i}^{c} (\{N_{i}^{c}\},\{N_{\beta}^{s}\}) +N_{i}^{c} -1 =
G_{i}^{0}+ N_{i}^{c} -1 -
 \sum_{j} g_{ij}^{cc} N_{j}^{c} - \sum_{\alpha} g_{i\alpha}^{cs}N_{\alpha}^{s},
\end{equation}
\begin{equation}
\tilde{D}_{\alpha}^{s} (\{N_{i}^{c}\},\{N_{\beta}^{s}\})=
D_{\alpha}^{s}(\{N_{i}^{c}\},\{N_{\beta}^{s}\}) +N_{\alpha}^{s} -1=
G_{\alpha}^{0} +N_{\alpha}^{s} -1-
\sum_{j} g_{\alpha j}^{sc} N_{j}^{c} - \sum_{\beta} g_{\alpha \beta}^{ss}
N_{\beta}^{s}.
\end{equation}
Rewriting these equations for each of the intervals $\Delta k$ and $\Delta
\Lambda$ we easily  find that in the $\Delta k \rightarrow 0$
and $\Delta \Lambda \rightarrow 0$ limits these four types of
statistical interactions reduce to the following form
\begin{equation}
g^{cc}(k,k') = \delta(k-k'),
\end{equation}
\begin{equation}
g^{cs}(k,\Lambda)=g^{sc}(\Lambda,k)= g^{ss}(\Lambda, \Lambda ') = 0.
\end{equation}
Thus, the vanishing $g$ functions in (21) simplify the expression (17) for the
total number of available configurations, which is then
\begin{equation}
\Omega = \prod_{i=1}^{N} \frac{ (G_{i}^{0})!}{(N_{i}^{c})!
(G_{i}^{0} -N_{i}^{c})!} \prod_{\alpha=1}^{M}
\frac{(G_{\alpha}^{0} + N_{\alpha}^{s} -1)!}{(N_{\alpha}^{s})!(G_{\alpha}^{0}-
1)!}.
\end{equation}
The statistical weight $\Omega$ factorizes  into the holon
$(\Omega^c)$ and the spinon $(\Omega^s)$ parts.
This, once again, expresses the fact that the spin and the charge degrees of
freedom decouple in the $U\rightarrow \infty$ limit \cite{ogata}.
As a consequence,
the entropy of the system is a sum of the two parts $S=S^c+S^s=
k_B \ln \Omega^c + k_B \ln \Omega^s$,
where the corresponding expressions calculated per particle are
\begin{equation}
S^c = - k_B \frac{1}{N_a} \sum_{i=1}^{N} \left[ n_i^c \ln n_i^c +
(1-n_i^c) \ln (1-n_i^c) \right],
\end{equation}
and
\begin{equation}
S^s = - k_B \frac{1}{N_a} \sum_{\alpha=1}^{M} \left[ n_i^s \ln n_i^s -
(1+n_i^s) \ln (1+n_i^s) \right],
\end{equation}
where $N_a$ is the number of atomic sites.

We recognize immediately that the holon contribution to the system entropy
coincides with that for spinless fermions, whereas the spinon contribution
reduces to localized-spin moments ($k_B \ln 2$) in the
Mott-insulator limit and in the spin disordered phase, i.e. when
$n_{\alpha}^s =1$ and $M=N_a /2$.
In general, one may say that $S^c$ provides the entropy of charge excitations
(and vanishes in the Mott insulating limit $n_i^c=1$), whereas $S^s$
represent the spin part of the excitation spectrum.
This demonstrates again that the holon (charge) excitations are fermions
and the spinon (spin) excitations are bosons.
In the $U \rightarrow \infty$ limit considered here the Heisenberg coupling
constant $(J=4t^2/U)$ vanishes and the spin wave excitations do not interact
with each other \cite{mat}.
In other words, they are dispersionless bosons.
Also, the charge excitations are spinless fermions.
The total entropy of the system reduces  in the Mott insulating
spin-disordered limit to $S=k_B \ln 2$.
This value is different from that for the Fermi liquid in the high-temperature
limit, which is $2 k_B \ln 2$.
This difference confirms on a statistical ground the inapplicability  of the
Fermi liquid concept to the Hubbard model in one dimension.

\section{Statistical interaction for the spin liquid}

In this section we derive the statistical interaction for the so-called
statistical spin liquid.
This concept  was introduced in
\cite{s1} to describe the thermodynamic properties of  strongly
interacting electrons.
The basic assumption in this approach is to exclude the doubly occupied
configurations of electrons with spin up and down not only in
real space but also in  reciprocal space (with given ${\bf k}$).
This assumption leads to a novel class of universality for
electron liquids.
Its thermodynamics in the normal, magnetic and superconducting states
were examined in the series of papers \cite{s1,nasze}.
A justification of this approach has as  its origin  in the concept of the
singularity in the forward scattering amplitude due to interparticle
interactions.
Namely, it was noted by Anderson \cite{and} and  by Kveshchenko \cite{khv}
that in two spatial
dimensions this amplitude may be divergent either due to the
Hubbard on-site repulsion, or due to the long-distance current-current
interaction mediated by the transfer gauge fields.
With the assumption that in those situations
 the wave vector is still a good
quantum number, one can write down
the phenomenological Hamiltonian describing
such liquid in the form
\begin{equation}
H=\sum_{\bf k \sigma} (\epsilon_{\bf k} - \sigma h)N_{\bf k \sigma}
+U_s \sum_{\bf k} N_{\bf k \uparrow}N_{\bf k \downarrow}.
\end{equation}
In this model $\epsilon_{\bf k}$ is the dispersion relation for the
particles with the wave vector ${\bf k}$ moving in the applied external
magnetic field $h$, $N_{\bf k \sigma}$ is the number of electrons in the state
$|\bf k \sigma>$, and $\sigma = \pm 1$ is the projected spin direction.
The number of double occupancies in given ${\bf k}$ state
is $N_{{\bf k} d}=N_{\bf k \uparrow}N_{\bf k \downarrow}$.
The    nonvanishing $N_{{\bf k} d}$ causes an
increase of the system energy  by
$U_s > 0 $ for each doubly occupied ${\bf k}$ state.
Finally, we will put $U_s \rightarrow \infty$ because
this model  is to represent the situation with the
singular forward scattering amplitude.
It turns out that this exactly solvable model \cite{s2} belongs to the class
of models with  Haldane's fractional statistics, as shown below.
In this case the statistical interaction is proportional
 to $\delta_{\bf k k'}$ in
 ${\bf k}$ space, but is a nondiagonal matrix in the extended spin space.

To prove this we define the total size of the Hilbert space of the
many-particle
states determined by the number of physically inequivalent configurations
\begin{equation}
\Omega = \prod_{\bf k}
\frac{
( D_{\bf k \uparrow} + N_{\bf k \uparrow} - N_{ {\bf k} d} -1 )!
}{
(N_ {\bf k \uparrow} - N_{ {\bf k} d } )!(D_{ \bf k \uparrow }  -1)!
}
\frac{
( D_{ \bf k \downarrow } + N_{ \bf k \downarrow } - N_{ {\bf k} d } -1 )!
}{
( N_{ \bf k \downarrow } - N_{ {\bf k} d })!( D_{ \bf k \downarrow }-1)!
}
\frac{
( D_{ {\bf k} d } + N_{ {\bf k } d } -1 )!
}{
( N_{ {\bf k} d })!( D_{ {\bf k} d }-1)!
}.
\end{equation}
Due to the local nature of the
interaction in  ${\bf k}$ space we must treat
separately the singly occupied states as distinct
 from those with double occupancy
in  reciprocal space.
Then, the statistical weight $\Omega$ expresses the possible ways of
distributing $N_{\bf k \sigma} - N_{{\bf k} d}$ quasiparticles over $D_{\bf k
\sigma}$ states and $N_{{\bf k} d }$ quasiparticles over the $D_{{\bf k} d}$
states.
In general, the dimension of the one particle Hilbert space
for the singly $(D_{\bf k \sigma})$ and the double occupied $(D_{{\bf k}d})$
states is the function of the number of other quasiparticles
$\{{N_{\bf k \sigma}}\}$ and $\{{N_{{\bf k}d}}\}$ \cite{s1,s2} i.e.,
\begin{equation}
D_{{\bf k} \alpha}(\{N_{{\bf k} \beta}\}) = G_{{\bf k}}^{0} -
\sum_{\beta} g_{\alpha, \beta} ({\bf k},{\bf k}')(N_{{\bf k}' \beta}
-\delta_{\alpha \beta} \delta_{{\bf k} {\bf k}'}),
\end{equation}
where $\alpha$ and $\beta$ label the configurations ${\uparrow,\downarrow,d}$;
these states define the extended spin space.
Note that in difference with Eqs.(15) and (16) we define here the boundary
conditions via the relations
$ D_{{\bf k} \alpha}(\{ N_{{\bf k} \beta}
= \delta_{\alpha \beta} \delta_{{\bf k k'}}\})
=G^{0}_{{\bf k}}$, i.e. the maximal dimension of the single-particle Hilbert
space is defined for an occupied configuration in each category, not for
an empty one.
This new conditions are equivalent to the form appearing in Eqs. (15) and (16),
in the thermodynamic limit.
Since, the Hamiltonian (25) does not mix different momenta,
we find the general solution for $g_{\alpha \beta}({\bf k},{\bf k}')$
in the form
\begin{equation}
g_{\alpha \beta}({\bf k},{\bf k}')=\delta_{{\bf k}{\bf k}'} \otimes
g_{\alpha \beta}.
\end{equation}
Hence,  the statistical interaction is diagonal in ${\bf k}$ space
for this model.
Next, to get the exact solution of the Hamiltonian (25) we choose
\begin{equation}
\renewcommand{\arraystretch}{1.5}
\everymath={\displaystyle}
g_{\alpha \beta}=
\left(
\begin{array}{lcr}
1 &1&-1\\
0 &1&
 0 \\
0&0&
1  \
\end{array}
\right),
\end{equation}
and substitute (29) and (28) into (27)
we find that the total size of the many-particle
Hilbert space is given by
\begin{equation}
\Omega = \prod_{\bf k}
 \frac{(G_{\bf k}^{0})!}{
(N_{\bf k\uparrow}-N_{{\bf k}d})! (N_{\bf k \downarrow} - N_{{\bf k}d})!
(N_{{\bf k}d})! (G_{\bf k}^{0} - N_{\bf k \uparrow} - N_{\bf k \downarrow}
+N_{{\bf k} d})!}.
\end{equation}
This result is exactly the same as that obtained in Ref.\cite{s1}
(cf. Appendix B).
Therefore, we conclude that this
model also belongs to the class of models with Haldane's  statistics.
In this case, the change in the distribution functions are not due to
the phase shift between  different momenta but rather
due to the mutual (dynamic) interaction between quasiparticles with the same
${\bf k}$ but different spin.
The interaction  pushes some  of the
states upward in energy,
 leading to the following form of momentum distribution functions

\begin{equation}
\frac{ N_{\bf k \sigma}-N_{{\bf k} d} }{ G_{\bf k}^{0} } =
\frac{ e^{\beta U_s} e^{\beta (\epsilon_{\bf k}-\mu)} \cosh (\beta h)}
{1+e^{\beta U_s}e^{\beta (\epsilon_{\bf k}-\mu)}
[e^{\beta (\epsilon_{\bf k}-\mu)}+2 \cosh(\beta h)]}
[1 +\sigma \tanh(\beta h)],
\end{equation}
\begin{equation}
\frac{ N_{{\bf k} d} }{ G_{\bf k}^{0} }=\frac{1}
{1+e^{\beta U_s}e^{\beta (\epsilon_{\bf k}-\mu)}
[e^{\beta (\epsilon_{\bf k}-\mu)}+2 \cosh(\beta h)]},
\end{equation}
which are easily  obtained by
 minimizing the thermodynamic potential with respect
to
$N_{\bf k \sigma}$ and $N_{{\bf k} d}$ separately \cite{s2}.
It is easy to show that those distributions evolve from the Fermi-Dirac
function to the statistical spin liquid distribution when
the $U_s$ changes from zero to  infinity \cite{s1}.

The limit $U_s\rightarrow \infty$ represents
the physical situation
in which $N_{{\bf k} d} \equiv0$.
In other words, there are no double occupancies  in  ${\bf k}$
space.
All  states are singly occupied by the quasiparticles with either spin up
or down, or empty.
In this limit,
the statistical interaction (28) reduces to the $2\times 2$ matrix form
\begin{equation}
\renewcommand{\arraystretch}{1.5}
\everymath={\displaystyle}
g_{\sigma \sigma'}({\bf k},{\bf k}')=
\delta_{{\bf k},{\bf k}'}\otimes
\left(
\begin{array}{cc}
1 &
1 \\
0 &
 1 \
\end{array}
\right)
{}.
\end{equation}
Then, the statistical weight is \cite{s1}
\begin{equation}
\Omega=\prod_{\bf k} \frac{(G_{\bf k}^{0})!}{
(N_{\bf k \sigma})!(G_{\bf k}^{0} -N_{\bf k \sigma} -N_{\bf k \bar{\sigma}})!}.
\end{equation}
Such liquid is called the {\bf statistical spin liquid}.
This is a novel class of quantum liquids,
similar, in some respect, to the Bethe-Luttinger liquid discussed above.
For example,
because of the mutual interaction between spin up and down particles,
 half of the total number of states ($2N_a$)
are pushed out of the physical space in the $U_s\rightarrow \infty$ limit.
Therefore, the entropy of the statistical spin liquid in the high-temperature
limit is the same as in the case of the Hubbard chain with the
infinite interaction
because the entropy of the system in this temperature limit is given
in terms of the degeneracy of the state only \cite{ch}.
In particular, the entropy in the statistical spin liquid for $N=N_a$ equals
$k_B N \ln 2$, which is the correct value for a Mott insulator.
Also, the high-temperature value of a thermopower is the same for
both liquids \cite{ch}.
It was also shown \cite{s3}
that the magnetization of statistical spin liquid has the
same form as that of the Hubbard chain with the infinite repulsion, i.e. that
for localized moments \cite{sok}.
However, the direct consequence of the statistical interaction is also a
 breakdown
of the Luttinger theorem:
the volume enclosed by the Fermi surface is twice that  for the Fermi
liquid.
This arises because of the differences in the microscopic character of the
single-particle excitations in these two liquids.

\section{Conclusions}

In this paper we  considered statistical properties of the two model
 system: The Hubbard chain with infinite repulsion and the statistical spin
liquid.
We  determined the form of the Haldane statistical interaction in
each case.
In the one-dimensional Hubbard model a novel distribution function emerges
due to the presence of the phase shift between pairs of states with
different rapidities $\{\Lambda_{\alpha}\}$ and $\{k_i\}$.
In the $U\rightarrow \infty$ limit, when all bound states are excluded,
the charge excitations (holons) behave statistically as fermions, and
the spin excitations (spinons) behave as bosons.
The holon have a simple energy dispersion $\epsilon_k$ coinciding with
the bare band energy, whereas the spinons are dispersionless.
An equivalent alternative approach \cite{car2} is based on the fermionic
representation of both the holon and the spinon degrees of freedom;
in that approach the spinons and the holons acquire complicated forms of the
effective dispersion relation.
The latter approach allows for a generalization of the treatment of the
Hubbard chain for finite $U$, the limit unavailable with our present
analysis.
Nonetheless, our analysis is also  applicable to other models when the
separation into the charge and the spin degrees of freedom occurs (cf. the
Kondo problem \cite{andr}).

In the statistical spin liquid case the mutual interaction between
quasiparticles with the same ${\bf k}$  leads to the
exclusion of the double
occupied configurations in reciprocal space.
In that case the statistical interaction is  diagonal in
${\bf k}$
 but has a nondiagonal structure in  spin space.
 In that case the statistical distribution changes with growing
 interaction from the Fermi-Dirac form to the spin liquid form \cite{s1}.

It is interesting that those two models of  particles with the internal
symmetry can be classified as the models with the fractional statistics
in the Haldane sense.
By contrast to  the case with scalar particles \cite{be},
the distribution functions in the present situation take the form of either
holon-spinon
 or the spin liquid distributions.
Those possibilities arise only when the particles have some internal
symmetry (spin,colour).
One may also say that in those cases the statistical interactions have
a tensorial character
in  space in which the Hamiltonian is diagonal.\\
\vspace{1cm}

The work was  supported
by the Committee of Scientific Research (KBN) of Poland,
Grants Nos. 2 P302 093 05 and 2 P302 171 06.
The authors are also grateful to the Midwest Superconductivity Consortium
(MISCON) of U.S.A. for the support through Grant No. DE-FG 02-90 ER 45427.


\begin{references}
\bibitem{hei}
e.g. W.Heisenberg, "The physical principles of the quantum theory",
(Dover Publications, INC (1949)).

\bibitem{l} J.M.Leinaas, J.Myrheim, Nuovo Cimento {\bf 37}, 1 (1977).

\bibitem{hal} F.D.M.Haldane, Phys.Rev.Lett. {\bf 67}, 937 (1991).

\bibitem{ms} M.V.N.Murthy, R.S.Shankar, Phys.Rev.Lett. {\bf 72}, 3629 (1994).

\bibitem{wu} Y.S.Wu, preprint (February 1994).

\bibitem{be} D.Bernard, Y.S.Wu, preprint (March 1994).

\bibitem{s1} J.Spa{\l}ek, W.W\'ojcik, Phys.Rev. {\bf B37}, 1532 (1988).

\bibitem{hub} J.Hubbard, Pros.Roy.Soc. {\bf A276}, 238 (1963).

\bibitem{lieb} E.H.Lieb, F.Y.Wu, Phys.Rev.Lett {\bf 20}, 1445 (1968).

\bibitem{ogata} M.Ogata, H.Shiba, Phys.Rev.{\bf B 41}, 2326 (1990).

\bibitem{sok} J.B.Sokoloff, Phys.Rev. {\bf B2}, 779 (1970).

\bibitem{got} Note that the proper normalization of the total density
 of states $(\rho_t^s)$ needs the following limiting procedure
 $\lim_{L \rightarrow \infty} (L/2 \pi) \delta_{\Lambda, \Lambda '} =
 \delta(\Lambda - \Lambda ')$; see e.g. K.Gottfried, "Quantum Mechanics"
 (W.A.Benjamin, inc., New York (1966)), vol.1, p.59.

\bibitem{car} J.Carmelo, D.Baeriswyl, Phys.Rev. {\bf B37}, 7541 (1988).

\bibitem{mat} D.C.Mattis, "The theory of magnetism" (Springer Verlag,
Berlin (1981)), vol.1, chapter 5.

\bibitem{nasze} J.Spa{\l}ek et al, Physica Scripta {\bf T49}, 206 (1993);\\
K.Byczuk, J.Spa{\l}ek, Acta Phys.Polonica {\bf A85}, 337 (1994);\\
J.Spa{\l}ek, Acta Phys.Polonica {\bf A85}, 39 (1994);\\
J.Spa{\l}ek, W.W\'ojcik, Acta Phys.Polonica {\bf A85}, 357 (1994).

\bibitem{and} P.W.Anderson, Phys.Rev.Lett. {\bf 65}, 2306 (1991); {\bf 66},
3226 (1991).

\bibitem{khv} D.V.Khveshchenko, Phys.Rev. {\bf B47}, 3446 (1993).

\bibitem{s2} J.Spa{\l}ek, Physica {\bf B 163}, 621 (1990).

\bibitem{s3} J.Spa{\l}ek, Phys.Rev. {\bf B40}, 5180 (1989).

\bibitem{car2} J.Carmelo, A.A.Ovchinnikov, J.Phys.:Cond.Matt. {\bf 3},
757 (1991).

\bibitem{ch} P.M.Chaikin, G.Beni, Phys.Rev.{\bf B 13}, 647 (1976).

\bibitem{andr} N.Andrei et al, Rev.Mod.Phys. {\bf 55}, 331 (1983).

\end{references}
\end{document}